# Reply to "Comment on 'Approximation for a Coulomb-Volkov solution in strong fields' "

Jarosław H. Bauer*

Katedra Fizyki Teoretycznej Uniwersytetu Łódzkiego,
Ul. Pomorska 149/153, 90-236 Łódź, Poland

It is shown that the criticism of Voitkiv [Phys. Rev. A **74**, 027403 (2006)] is based on one assumption, which is false for an ionized electron, initially bound by the Coulomb potential, in a strong circularly polarized laser field. A certain new Coulomb correction to the Volkov wavefunction is proposed.

___________________________

*E-mail: bauer@uni.lodz.pl

In the recent Comment [1] the paper of Reiss and Krainov [2] entitled "Approximation for a Coulomb-Volkov solution in strong fields" has been criticized. The author of Ref. [1] claims that the derivation of the Coulomb correction [2] to the well-known Volkov wavefunction [3,4] "…contains an error caused by confusing the space coordinates before and after the Kramers-Henneberger transformation" (KH) [5,6]. (It is proper to add that Reiss and Krainov do not agree with Voitkiv's criticism [7].) This modified Volkov wavefunction [see Eq. (10) of Ref. [2]] or its generalizations have been used by many authors to calculate photoionization rates in a strong circularly [2,8,9] or linearly [10,11] polarized laser field. Although one might doubt about the justification of such Coulomb corrections, it has been very well verified that they always increase ionization rates calculated in the velocity gauge [2,8,9,11], what causes a better agreement of the so-called strong-field approximation (SFA) [12] for atoms with other experimental or theoretical results.

In the text below Eq. (6) of Ref. [1] we can read that "…it is certainly true that the main contribution to the transition matrix element is given by the region of small electron-nucleus distances where the atomic ground state is located…". Further, below Eq. (7), we can read that "…in the region $r \leq \propto 1/Z$, which is the only region where the state $\Psi(\vec{r},t)$ may have a substantial overlap with the initial (ground) atomic state and which is thus most important for the transition matrix element…". ($r = |\vec{r}|$ denotes the distance of the bound electron from the nucleus of the total charge $Z$; here we use atomic units and the same notation as in Refs. [1,2].) It is probably the $\exp(-Zr)$ form of the initial state wavefunction that brought the author of Ref. [1] to this conclusion. But certainly there is no proof given in Ref. [1] for such statement, which has become the basic assumption of all Voitkiv's criticism [see the paragraph containing Eq. (7) in Ref. [1]]. Although a similar statement one may found also in Ref. [2], one does not have to follow the reasoning of Reiss and Krainov (based on the KH frame) to derive their result [2,7]. In what follows we will show, using only the laboratory frame of reference, that for ionization in a circularly polarized strong laser field the above mentioned region (whose contribution to the transition matrix element is significant) is much greater. In fact, the region $r \leq 1/Z$, is indeed very important for the transition matrix element, but this is not the only important region. It appears that the intuition (probably) coming from perturbative laser fields can be misleading in strong fields.

In his considerations the author of Ref. [1] debates the properties of a wavefunction which is the exact solution of the following time-dependent Schrödinger equation (TDSE) in the dipole approximation

$$i\frac{\partial}{\partial t}\Psi(\vec{r},t) = \left[\frac{1}{2}\left(-i\vec{\nabla} + \frac{1}{c}\vec{A}(t)\right)^2 - \frac{Z}{r}\right]\Psi(\vec{r},t) . \qquad (1)$$

In the $S$ matrix theory of photoionization [12] one uses stationary solutions of the TDSE with positive energy $E = \vec{p}^{\,2}/2 > 0$ and $\vec{p}$ - the asymptotic (when $r \to \infty$) momentum as a parameter. Since one cannot solve Eq. (1) analytically, it is rather difficult to take something for granted in this general case, particularly if neither the laser field, nor the Coulomb one can be treated perturbatively. However, for sufficiently strong laser field one can approximate the solution of Eq. (1) by the solution of the following equation

$$i\frac{\partial}{\partial t}\Psi^V(\vec{r},t) = \frac{1}{2}\left(-i\vec{\nabla} + \frac{1}{c}\vec{A}(t)\right)^2 \Psi^V(\vec{r},t) . \qquad (2)$$

This solution is the well-known nonrelativistic Volkov wavefunction (where $\vec{p}$ is a parameter)

$$\Psi^V_{\vec{p}}(\vec{r},t) = \frac{1}{(2\pi)^{3/2}} \exp\left[i\vec{p}\vec{r} - \frac{i}{2}\int_{-\infty}^{t}\left(\vec{p} + \frac{1}{c}\vec{A}(\tau)\right)^2 d\tau\right] . \qquad (3)$$

Substituting Eq. (3) into Eq. (1) one obtains

$$\frac{1}{2}\left(\vec{p}+\frac{1}{c}\vec{A}(t)\right)^2 \Psi_{\vec{p}}^V(\vec{r},t) \cong \left[\frac{1}{2}\left(\vec{p}+\frac{1}{c}\vec{A}(t)\right)^2 - \frac{Z}{r}\right]\Psi_{\vec{p}}^V(\vec{r},t) . \qquad (4)$$

One should expect, looking at Eq. (4), that this approximate equality improves when either $\vec{p}$ or the amplitude of $\vec{A}(t)$ increases. In photoelectron energy spectra, due to selection rules for angular momentum, one obtains the greatest $n$-photon ionization rates for the certain values of the final kinetic energy $E = p^2/2$ of the outgoing electron. These values depend on the kind of polarization of the laser field and on its intensity. For linear polarization the low-energy electrons always dominate, but for circular polarization electrons with much higher energy $E = p^2/2 \cong U_P = I/(4\omega^2)$ usually dominate ($U_P$ is the ponderomotive potential and $I, \omega$ are the laser intensity and frequency) [13]. Therefore for the 1$s$ H atom ($Z=1$), particularly for circular polarization and high intensity of the laser field, for most ionized electrons Eq. (4) is very well satisfied. In other words, these electrons appear well in the continuum of ionized states and the Volkov wavefunction is a very good approximation to the exact solution of Eq. (1). Much the same, the Coulomb-Volkov wavefunction derived by Reiss and Krainov [2]

$$\Psi_{\vec{p}}^{CV}(\vec{r},t) = \Psi_{\vec{p}}^V(\vec{r},t)\exp\left(\frac{iZ}{\alpha_0}t\right) = \frac{1}{(2\pi)^{3/2}}\exp\left[i\vec{p}\vec{r} - \frac{i}{2}\int_{-\infty}^{t}\left(\vec{p}+\frac{1}{c}\vec{A}(\tau)\right)^2 d\tau + \frac{iZ}{\alpha_0}t\right]$$

(5)

(where $\alpha_0$ is the radius of the circular classical motion of a free electron in a circularly polarized laser field) is a very good approximation to the exact solution of Eq. (1) too. One can also derive Eq. (5) using nonperturbative path-integral approximations [14,15], without utilizing the KH frame. Substituting Eq. (5) into Eq. (1) one obtains

$$\left[\frac{1}{2}\left(\vec{p}+\frac{1}{c}\vec{A}(t)\right)^2-\frac{Z}{\alpha_0}\right]\Psi_{\vec{p}}^{CV}(\vec{r},t)\cong\left[\frac{1}{2}\left(\vec{p}+\frac{1}{c}\vec{A}(t)\right)^2-\frac{Z}{r}\right]\Psi_{\vec{p}}^{CV}(\vec{r},t)\ . \qquad (6)$$

The improvement due to the appearance of the term $-Z/\alpha_0$ on the left-hand side in this approximate equality will be substantial only for large $r$, because one assumes that $\alpha_0 \gg 1/Z$ [2]. However, although the Coulomb correction $-Z/\alpha_0$ may be quite small (due to large $\alpha_0$), it can give significant increase of the calculated ionization rate (see, for example, Fig. 2 of Ref. [2] or Figs. 1, 2 of the second Ref. [9]).

Let us consider the influence of the distance of ionized electron from the center of an atom (or ion) on the ionization probability. Within the limits of the $S$ matrix theory of photoionization we can do it in the following way. The general analytical expressions for ionization rate in the SFA (and spherically symmetric initial state) are

$$\Gamma_{cir}^{SFA}=(2\pi)^2\sum_{n=n_0}^{\infty}p_n\left[\tilde{\Phi}_i(p_n)\left(\frac{p_n^2}{2}+E_B\right)\right]^2\int_0^{\pi}d\vartheta\sin\vartheta\, J_n^2\left(\sqrt{\frac{2z}{\omega}}p_n\sin\vartheta\right), \qquad (7a)$$

$$\Gamma_{lin}^{SFA}=(2\pi)^2\sum_{n=n_0}^{\infty}p_n\left[\tilde{\Phi}_i(p_n)\left(\frac{p_n^2}{2}+E_B\right)\right]^2\int_0^{\pi}d\vartheta\sin\vartheta\, J_n^2\left(2\sqrt{\frac{z}{\omega}}p_n\cos\vartheta,-\frac{z}{2}\right) \qquad (7b)$$

for circular and linear polarization respectively, where $p_n^2/2=(n-z)\omega-E_B$ is the final kinetic energy of the outgoing electron and its binding energy in the initial state is $E_B$. [In Eqs. (7) one does not necessary assume that $E_B=Z^2/2$.] If this state is the 1$s$ wavefunction of H atom in the momentum space (normalized to unity in the entire space) $\tilde{\Phi}_i(p)$, then $E_B=Z^2/2=0.5$ $a.u.$ [We refer the reader to our recent paper [16] and to Ref. [12] for more details regarding Eqs. (7).] The expressions (7) have been obtained after integration over all possible positions $\vec{r}$ of the electron in space (when calculating the $S$ matrix element). This means that all possible electron distances from the center of an atom (or ion) $0\leq r\leq\infty$ contribute to ionization rates in Eqs. (7). But according to Ref. [1] only $r\leq\propto 1/Z$ would contribute. Therefore it would be interesting to check what is indeed the maximum distance $R$ of the electron from the center of an atom (or ion), which really contributes to the ionization rates in Eqs. (7). To this end we have derived expressions analogical to Eqs. (7), but with $0 \leq r \leq R$. Therefore instead of the initial state wavefunction $\Phi_i(r)$ we have taken

$$\Phi_R(r) = \Phi_i(r)(1 - \theta(r - R)) = \sqrt{\frac{Z^3}{\pi}} \exp(-Zr)(1 - \theta(r - R)), \qquad (8a)$$

[$\theta(x)$ is the Heaviside step function] which leads to

$$\tilde{\Phi}_R(p) = \int \frac{d^3 r}{(2\pi)^{3/2}} \Phi_R(r) \exp(-i\vec{p}\vec{r}) = \frac{\sqrt{2Z^3}}{\pi p (p^2 + Z^2)^2} \times$$
$$\times \{2Zp + e^{-ZR}[\sin pR(p^2 - p^2 ZR - Z^2 - Z^3 R) - \cos pR(p^3 R + 2pZ + pZ^2 R)]\} \qquad (8b)$$

[From Eq. (8b) one obtains the well-known expression for $\tilde{\Phi}_i(p)$ in the limit $R \to \infty$.] Replacing $E_B = Z^2/2$ by $E'_B = Z^2/2 - Z/\alpha_0$ in Eq. (7a) one obtains the so-called Coulomb corrected strong-field approximation (CSFA) [2]. In Fig. 1 we plot the CSFA ionization rate of the $1s$ H atom as a function of intensity of the laser field for $\omega = 0.074$ $a.u.$ ($\lambda = 616$ $nm$ - a typical optical wavelength) and a few different $R$'s [thus $\tilde{\Phi}_i(p)$ is replaced by Eq. (8b) in Eq. (7a)]. The range of intensities corresponds to $10$ $a.u. \leq \alpha_0 \leq 100$ $a.u.$ in Fig. 1. Let us have a look at consecutive curves on this graph. It is obvious that certainly neither $R = 2$ $a.u.$ nor $R = 4$ $a.u.$ can properly describe ionization rates for all intensities shown here. We have also checked that even for a few intensities, where ionization rates are equal with those of $R = \infty$ (see the intersection points of the $R = 2$ $a.u.$ and $R = 4$ $a.u.$ curves with the $R = \infty$ one in Fig. 1) the photoelectron energy spectra are significantly deformed. One needs at least about $R = 6$ $a.u.$ to properly reproduce the $R = \infty$ result for the laser field parameters shown in Fig. 1. Moreover the limiting value of $R$, which reasonable describes ionization rate, grows with increasing intensity. What is even more interesting, in spite of the fact that the wavefuction (8a) is normalized to less than unity (for example, when $R = 2$ a.u. one obtains $\int d^3r |\Phi_R(r)|^2 \cong 0.762$), for some intensities and for finite $R$ one obtains much greater ionization rate than the true ($R = \infty$) CSFA result. It appears that the quantum-mechanical interference effect plays a very important role in the strong-field photoionization. For the highest intensities shown in Fig. 1 it is the destructive interference of different possible space positions $\vec{r}$ of ionized electron, roughly from $0 \le r \le 6$ a.u., that produces the true ($R = \infty$) CSFA result.

What is the effect of finite $R$ for much higher intensities of the laser field? We show this for the SFA [$E_B = Z^2/2$ in Eqs. (7)] in Figs. 2a and 2b for both circular and linear polarization respectively. The range of intensities corresponds to limitations of the nonrelativistic SFA in Figs. 2 ($1 \le z_1$ and $z_f \le 0.1$, where $z_1 = 2U_P/E_B$ and $z_f = 2U_P/c^2$; see Refs. [12,16] for more details). It appears that for circular polarization and laser fields strong enough $R = 6$ a.u. or even $R = 8$ a.u. are not sufficient to properly describe ionization rate in the velocity gauge $S$ matrix theory. In contrast, for linear polarization the assumption that only $r \le \infty 1$ a.u. contribute to ionization rate is quite well satisfied for the laser field parameters from Fig. 2b.

In principle, the SFA has been introduced for short-range potentials [12] and it should work better in this case than for the long-range Coulomb potential. Therefore let us also check the effect of finite $R$ for the zero-range potential for the same binding energy ($E_B = Z^2/2$) and laser field parameters as in the Coulomb potential case. On the analogy of Eqs. (8) we obtain

$$\Phi_{0R}(r) = \Phi_0(r)(1 - \theta(r - R)) = \sqrt{\frac{Z}{2\pi}} \frac{1}{r} \exp(-Zr)(1 - \theta(r - R)), \qquad (9a)$$

$$\tilde{\Phi}_{0R}(p) = \frac{\sqrt{Z}}{\pi(p^2 + Z^2)} \left[1 - e^{-ZR}\left(\frac{Z}{p}\sin pR + \cos pR\right)\right], \qquad (9b)$$

and we substitute Eq. (9b) into Eqs. (7). Also in this case, in the limit $R \to \infty$, one has the initial state wavefunction which is exact. Moreover, the SFA ionization rate for the zero-range potential is gauge-invariant, i.e. the length and velocity gauge results are identical [16,17]. In Figs. 3a and 3b we present ionization rate (for an electron bound by this potential) as a function of intensity for the circular and linear polarization of the laser field respectively. It appears that for the zero-range potential and both polarizations ionization always takes place roughly inside the sphere of the radius $R \cong 2$ $a.u.$. It follows from Fig. 3b that for linear polarization the radius of the sphere, where ionization takes place, decreases with increasing intensity. This is in good qualitative agreement with the evaluation given by Gribakin and Kuchiev in Ref. [17]. For example, for intensity $I = 0.01$, 0.1, and 1 $a.u.$ one obtains $R = 2.8$, 1.8 and 1.0 $a.u.$ respectively from Eq. (2) of Ref. [17]. [Note that in their paper the $H^-$ ion was considered with much smaller binding energy and therefore much larger $R$, but Eq. (1) of Ref. [17] is satisfied in our case, because $\omega = 0.074$ $a.u. << 0.5$ $a.u. = E_B$.]

It follows from Figs. 2 and 3 that the Coulomb potential case in circularly polarized strong laser field is exceptional in a way because photoionization takes place in much larger space than in the remaining three cases. Let us now treat the ionization rates for finite $R$ as a hint for finding a better Coulomb correction in the SFA ionization rate formula for circular polarization [Eq. (7a) with $R = \infty$]. Fig. 2a suggests that instead of the $-Z/\alpha_0$ Coulomb correction in Eq. (6) one could use $-Z/R_{eff}(\omega, I)$ to compensate partially the term $-Z/r$ on the right-hand side of this equation. The effective (or phenomenological) parameter $R_{eff}$ would be a certain function of the laser frequency and intensity. Therefore the Volkov wavefunction would be multiplied not by $\exp(iZt/\alpha_0)$, but rather by $\exp(iZt/R_{eff})$ in Eq. (5). As a result, in the SFA ionization rate formula the binding energy would be replaced by $E'_B = Z^2/2 - Z/R_{eff}$. In Fig. 4 we show such ionization rates (by two identical solid lines) as a function of intensity for the circularly polarized laser field for two different constant values $R_{eff} = 5$ $a.u.$ and $R_{eff} = 10$ $a.u.$. Roughly these values are suggested by Fig. 2a for $\omega = 0.074$ $a.u.$ as the limits between which the new Coulomb-corrected SFA ionization rate could run across. [For the lowest intensities shown in Fig. 2a $R_{eff} \cong 5$ $a.u.$, and for the highest intensities - $R_{eff} \cong 10$ $a.u.$. It has not been our aim here to find the best $R_{eff}(\omega, I)$ in general.] For comparison we also show the SFA and the CSFA ionization rates in Fig. 4. There are also some other theoretical calculations, which are valid for smaller intensities, but which have some common range of validity with the above mentioned various strong-field calculations. The Floquet calculations have been taken from Fig. 5 of Ref. [18]. The WKB Coulomb corrected KFR theory [19,20,16,21], in both gauges, has a high-intensity limit connected with existence of the Coulomb barrier and the critical laser field strength [18] in the 1$s$ H atom. The WKB-Reiss ionization rate has been calculated from Eq. (9a) of Ref. [21] and the WKB-Keldysh one from Eq. (32a) of Ref. [16]. One can easily observe that in Fig. 4, around $I = 10^{-2}\,a.u.$, the curve with $R_{eff} = 5\ a.u.$ is much closer to the Floquet and the WKB Coulomb corrected KFR results than the SFA and CSFA curves.

In conclusion, the main result of the present paper is revealing that photoionization takes place in much greater volume than a naive expectation ($r \leq\, \propto 1/Z$) would predict, if the following three conditions are simultaneously satisfied: (i) the binding potential is the (long-range) Coulomb one, (ii) the laser field is strong enough, and (iii) the laser field is circularly polarized. This is another explanation why the SFA should work much better for the circular polarization than for the linear one. For a given frequency and intensity of the laser field one could always try to find the approximate parameter $R_{eff}(\omega, I)$ resulting in a greater and more accurate ionization rate. It seems possible to modify at least some results of Refs. [8,9,11] by replacing $\alpha_0$ with suitably chosen $R_{eff}(\omega, I)$.

**ACKNOWLEDGMENTS**

The author is indebted to Professor Howard R. Reiss for his correspondence on the present subject. This paper has been supported by the University of Łódź.

**References.**

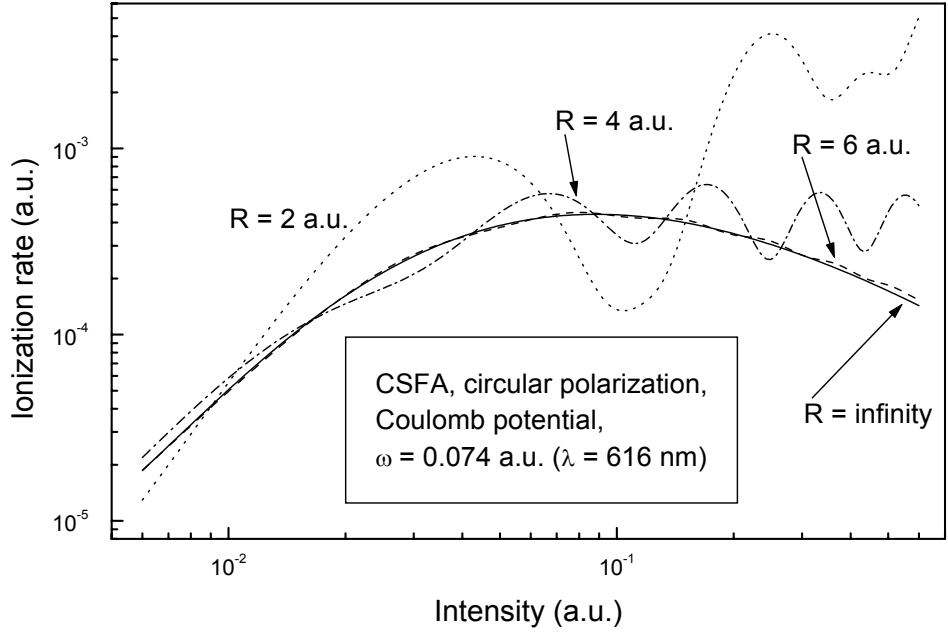

FIG. 1. The CSFA ionization rate (of the $1s$ H atom) as a function of intensity for $\omega = 0.074$ *a.u.* in the range of intensities corresponding to $10\ a.u. \leq \alpha_0 \leq 100\ a.u.$ for the circularly polarized laser field. The solid line ($R = \infty$) is the Reiss-Krainov result [2]. The other three ionization rates have been calculated assuming that only $0 \leq r \leq R$ contribute to the $S$-matrix element (respectively for $R = 2,\ 4,\ 6\ a.u.$).

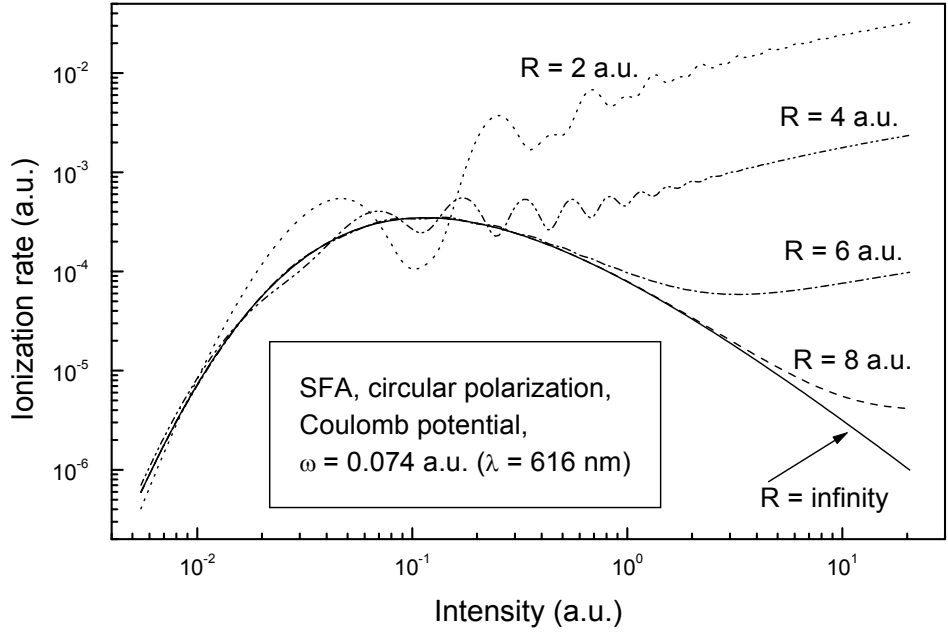

FIG. 2a. The SFA ionization rate (of the $1s$ H atom) as a function of intensity for $\omega = 0.074$ *a.u.* in the range of intensities corresponding to $1 \leq z_1$ and $z_f \leq 0.1$ for the circularly polarized laser field. The solid line ($R = \infty$) is the Reiss result [12]. The other four ionization rates have been calculated assuming that only $0 \leq r \leq R$ contribute to the $S$-matrix element (respectively for $R = 2, 4, 6, 8$ *a.u.*).

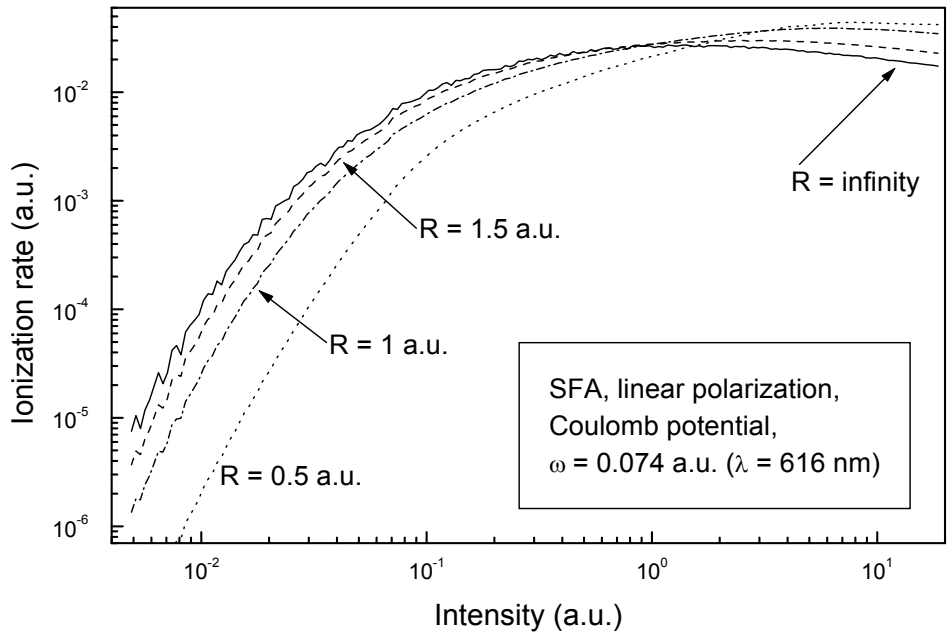

FIG. 2b. The SFA ionization rate (of the $1s$ H atom) as a function of intensity for $\omega = 0.074$ *a.u.* in the range of intensities corresponding to $1 \leq z_1$ and $z_f \leq 0.1$ for the linearly polarized laser field. The solid line ($R = \infty$) is the Reiss result [12]. The other three ionization rates have been calculated assuming that only $0 \leq r \leq R$ contribute to the $S$-matrix element (respectively for $R = 0.5, 1, 1.5$ *a.u.*).

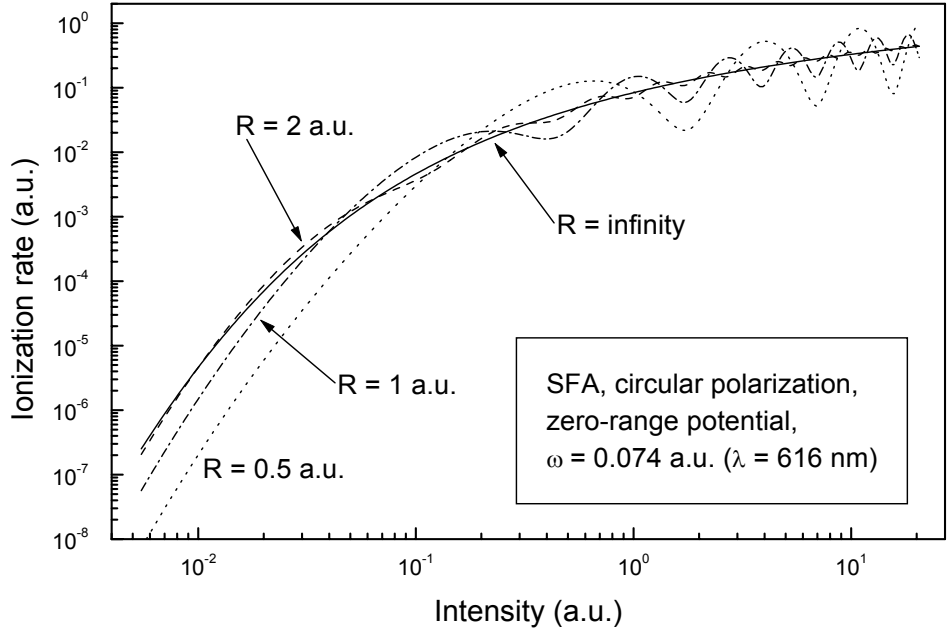

FIG. 3a. The SFA ionization rate (of the only bound state in the zero-range potential with $E_B = 0.5$ $a.u.$) as a function of intensity for $\omega = 0.074$ $a.u.$ in the range of intensities corresponding to $1 \leq z_1$ and $z_f \leq 0.1$ for the circularly polarized laser field. The solid line ($R = \infty$) is the result of Eq. (7a) [with the substitution of Eq. (9b) for $R \to \infty$]. The other three ionization rates have been calculated assuming that only $0 \leq r \leq R$ contribute to the $S$-matrix element (respectively for $R = 0.5, 1, 2$ $a.u.$).

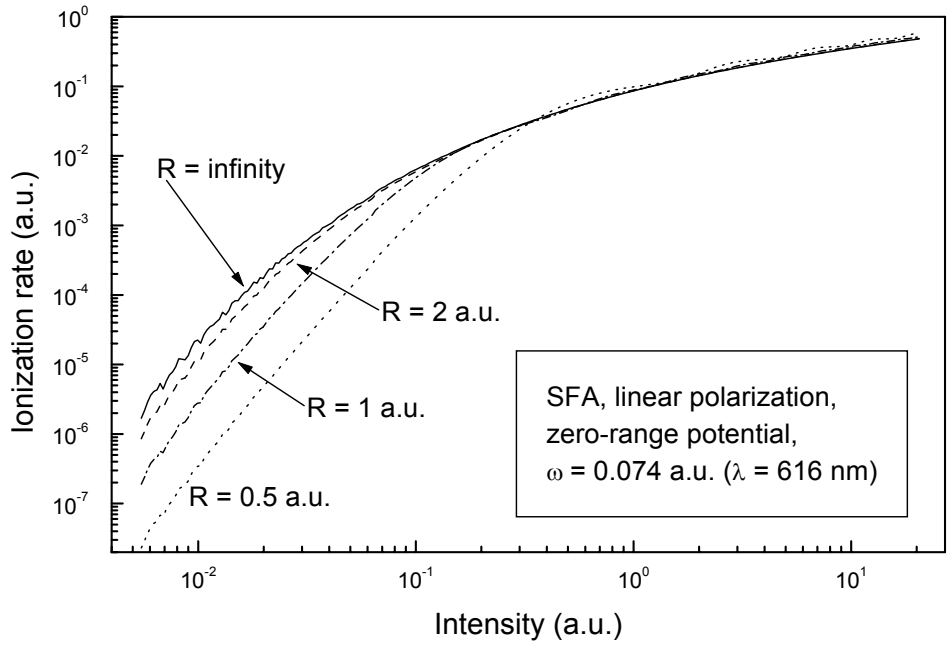

FIG. 3b. The SFA ionization rate (of the only bound state in the zero-range potential with $E_B = 0.5$ *a.u.*) as a function of intensity for $\omega = 0.074$ *a.u.* in the range of intensities corresponding to $1 \leq z_1$ and $z_f \leq 0.1$ for the linearly polarized laser field. The solid line ($R = \infty$) is the result of Eq. (7b) [with the substitution of Eq. (9b) for $R \to \infty$]. The other three ionization rates have been calculated assuming that only $0 \leq r \leq R$ contribute to the $S$-matrix element (respectively for $R = 0.5, 1, 2$ *a.u.*).

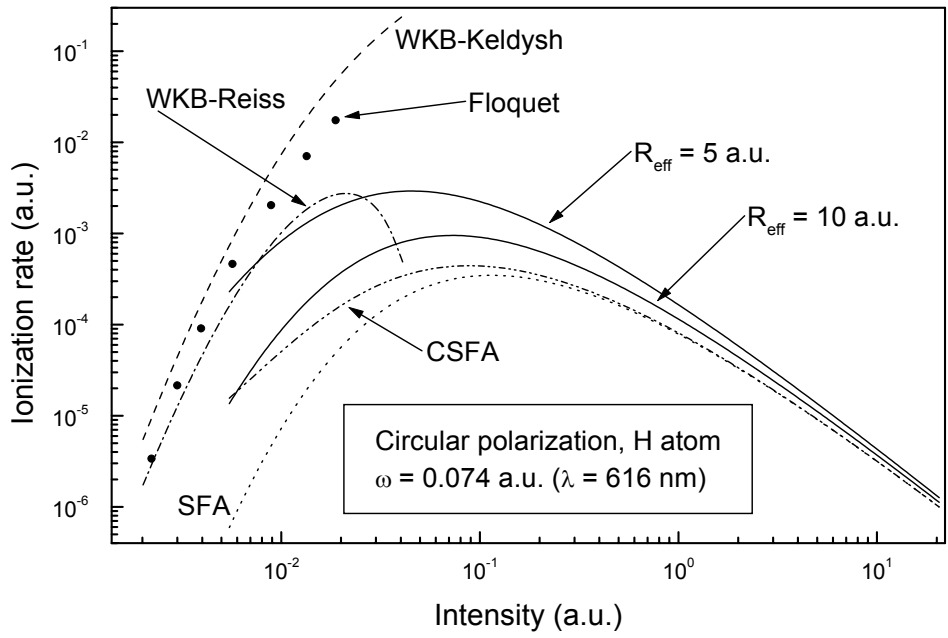

FIG. 4. Various theoretical ionization rates (of the 1*s* H atom) as a function of intensity for $\omega = 0.074$ *a.u.* and the circularly polarized laser field (see text for details).